\documentclass[runningheads]{svmult}

\usepackage{makeidx}   
\usepackage{graphicx}  
\usepackage{subeqnar}  
\usepackage{multicol}  
\usepackage{physprbb}  
\makeindex             

\begin{document}
\title*{A Million Element Integral Field Unit (MEIFU)}
\toctitle{A Million Element Integral Field Unit (MEIFU)}
%
%
\titlerunning{MEIFU}
%
\author{Simon Morris\inst{1}
\and Robert Content\inst{1}
\and Ray Sharples\inst{1}
\and Richard Bower\inst{1}
\and Roger Davies\inst{1}
\and Carlton Baugh\inst{1}}
\authorrunning{Simon Morris et al.}
%
%
\institute{Physics Department, Durham University, South Rd., Durham, DH1 3LE, 
ENGLAND}

\maketitle              

\begin{abstract}
We describe an instrument concept that will provide simultaneous
spectra for a million spatial samples on the sky. With the proposed
field of view and spectral resolution, it will be able to measure
redshifts and line strengths for around 2-4000 z$\sim$3-7 galaxies in a 16 night 
campaign. The main science driver is to obtain a complete census of the
star formation properties of galaxies with 2.5$<$z$<$6.7 as a function of
luminosity, environment, morphology and redshift. This survey
will also allow us to study the evolution of the 3 dimensional power
spectrum and test the connection between these early galaxies and
the large scale structure of the Lyman $\alpha$ forest. 

Additional science drivers include [OII] emitters with 0.13$<$z$<$1.52,
abundance gradients in nearby galaxies, searches for planetary nebulae in
nearby galaxy groups and clusters, lensed emission line objects, and
spectroscopic searches for objects with continuum breaks. We expect that
this instrument will make an ideal partner to the HST ACS, where ACS
morphology and broadband colour information can complement the MEIFU
spectroscopy and help remove any degeneracies in line identification. 

\end{abstract}

\section{Introduction}

In this contribution to the workshop on `Scientific Drivers for ESO Future 
VLT/VLTI Instrumentation', we would like to make a case for the study, 
construction and deployment of a wide-field, optical integral field unit (IFU). 
`Wide field' here means around 5 arcminutes diameter, with 0.3 arcsecond 
sampling, and `optical' means 420-940nm. As one might imagine, this implies a 
very large number of pixels in the final detector arrays (around 10$^{9}$), but 
we feel that modern CCD mosaics are well enough understood to allow this. It 
also implies a large number of spectrographs. We estimate that the instrument 
will need 27 spectrographs. 
However, these spectrographs can be kept relatively simple, and 
`mass production',  or a modular approach, can be used to reduce the cost and 
risk of the individual units. This also allows the instrument to be deployed in 
a 
phased manner.

In this contribution, we start by discussing the context for such an instrument. 
We then describe the 
instrument concept and make some preliminary Signal-to-Noise (S/N) estimates. We 
can then combine these with semi-analytic calculations for the expected observed 
Lyman $\alpha$ emission from galaxies as a function of redshift (including the 
effects of dust obscuration), to give predictions for the numbers of 
detections. We conclude with some of the other possible science cases for the 
instrument.

Table~\ref{tab1} lists some of the other facilities that will be demanding 
followup or complementary optical spectroscopy. As can be seen, many of these 
facilities have large fields of view (FOV).

\begin{table}[htbp]
\caption{The need for complimentarity with other facilities}
\begin{center}
\renewcommand{\arraystretch}{1.4}
\setlength\tabcolsep{5pt}
\begin{tabular}{ll}
\hline\noalign{\smallskip}
 Facility & Field of View \\
\noalign{\smallskip}
\hline
\noalign{\smallskip}
 Chandra 	& 17 x 17 arcmin  \\
 XMM 		& 30 arcmin diameter \\
 HST ACS 	& 3.3 x 3.3 arcmin \\
 SIRTF 		& 5.1 x 5.1 arcmin  \\
 NGST Imaging 	& 4 x 4 arcmin \\
 SCUBA 		& 2.3 arcmin diameter  \\
 ALMA 		& 12 arcsec diameter, but will mosaic fast \\
\hline
\end{tabular}
\end{center}
\label{tab1}
\end{table}

As demonstrated in Guy Monnet's introduction to the workshop, the VLT already 
has recognised this need, both with Multi-Object Spectrographs (MOS) with wide 
FOV, but very small filling factors within that FOV, and also with IFUs. We list 
some of these IFUs in table~\ref{tab2} in order to put the MEIFU proposal into 
context.

\begin{table}[htbp]
\caption{(Some) optical 8m IFUs (wide field option)}
\begin{center}
\renewcommand{\arraystretch}{1.4}
\setlength\tabcolsep{5pt}
\begin{tabular}{llll}
\hline\noalign{\smallskip}
Name	&	Telescope &	FOV (arcsec) &		Sampling \\
\noalign{\smallskip}
\hline
\noalign{\smallskip}
Flames		& VLT		& 11.5x7.3"	& 17x17 \\
GMOS		& Gemini	& 5x7"		& 32x32 \\
VIMOS		& VLT		& 54x54"	& 80x80 \\
FMOS		& Subaru	& 15x(5x5)	& 15x(15x15) \\
 & & & \\
MEIFU		& VLT		& 300x300"	& 1000x1000 \\
\hline
\end{tabular}
\end{center}
\label{tab2}
\end{table}

This will clearly require a 
quantum leap in IFU capabilities. We believe that this is possble, by using a 
combination 
of the lenslet developments from instruments such as OASIS and SAURON, together 
with a creative use of anamorphic magnification to improve the packing fraction 
on the detector. Both of these techniques are now well tested and understood.

\section{Instrument Concept}

\begin{figure}[htbp]
\begin{center}
\includegraphics[width=1.1\textwidth]{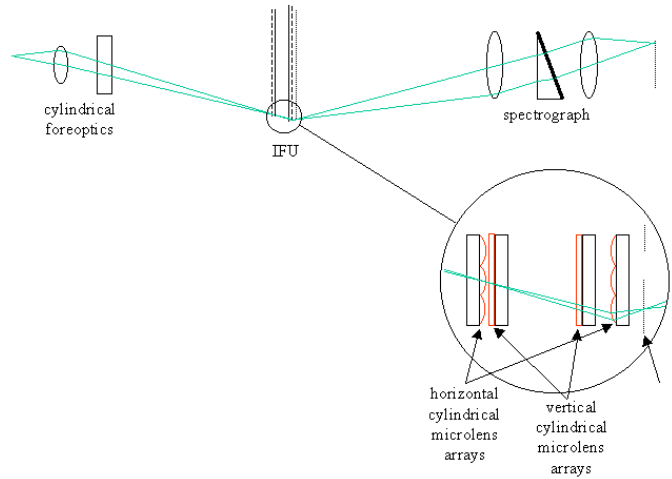}
\end{center}
\caption[]{Optical Principle: Fore-optics magnify the focal plane anamorphically 
(f/36 x f/195) onto the first (rectangular) lenslet array which acts as the 
image slicer. The first cylindrical microlens arrays divide the input focal 
plane into micro-slices (1.5x0.3 arcsec, 10x2 pixels). A second lenslet array 
anamorphically demagnifies the slices onto the input focal plane of the 
spectrograph and puts the pupil onto the grating.}
\label{eps1}
\end{figure}

\begin{figure}[htbp]
\begin{center}
\includegraphics[width=.9\textwidth]{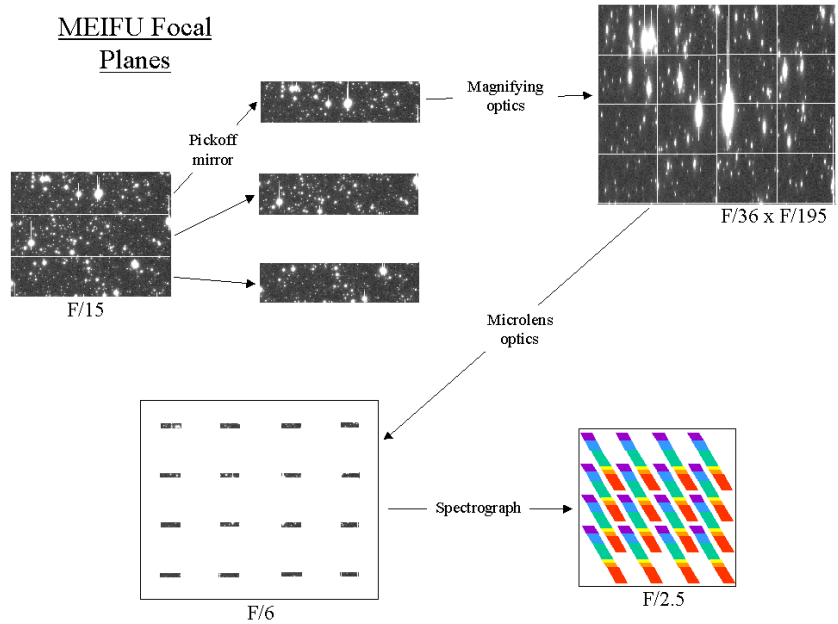}
\end{center}
\caption[]{MEIFU focal planes: Spectra are dispersed at an angle (13 deg) to 
avoid spectral overlap. The spectra are 200x12 pixels in the blue, and 675x12 
pixels in the red on the detector (0.15 arcsec per pixel). Inter-spectrum gaps 
are 26 pixels (spectral) and 3 pixels (spatial)}
\label{eps2}
\end{figure}

The basic instrumental parameters are listed below, while the concept is 
explained in figures~\ref{eps1} and \ref{eps2}. 

\begin{itemize}
\item Approximately a 5x5 arcmin field of view
\item 0.3x0.15 arcsec sampling (1000x2000 pixels)
\item cover the entire wavelength range 420-940 nm in 4 exposures (by rotating 
the spectrograph FOV through 90 degrees each time)
\item Spectral resolution (2 pixels) R$\sim$500 (420-750 nm), and 1500 (750-940 
nm). The larger resolution in the red is needed because of the strong 
atmospheric OH emission.
\item 27 spectrographs (12 covering 420-750nm, and 15 covering 750-940nm, each 
with 6kx6k CCD mosaic detectors)
\item With a change in the fore-optics one could change the instrumental plate 
scale and convert the instrument to something optimised for Multi-Conjugate AO.
\item The intention is to use this 
instrument with an optical derotator and to mount it on the floor of a Naysmith 
platform. 
\item Detailed mass and costs budgets have not been developed, but BOTE 
estimates suggest that we can fit into the VLT mass limit and also within our 
guesses as to the available funding for second generation VLT facility 
instruments.
\end{itemize}

We propose to use crossed cylindrical lenslet arrays made of fused silica and 
anti-reflection coated. The big advantage of this approach is that it is easy to 
make a lenslet array with different f-ratios in the x and y directions, i.e. it 
is naturally anamorphic.

Given the need for 27 spectrographs, reducing their cost and complexity is 
clearly crucial. Some of the key ideas we had to develop in order to do this 
were:
\begin{itemize}
\item de-magnifying the global pupil with the second lenslet arrays 
\item having non-parallel rays at Volume Phase Holographic (VPH) grating
\item only requiring 2 wavelength achromatism (helped by also only requiring a 
fairly small simultaneous wavelength coverage in any given spectrograph).
\item maximising the use of off-the-shelf lenses
\item splitting the field into 4 wavelength segments 420-510, 510-620, 620-750, 
750-940 nm, so spectrographs in each quadrant only cover one of these wavelength 
ranges.
\end{itemize}

\begin{figure}[htbp]
\begin{center}
\includegraphics[width=.9\textwidth]{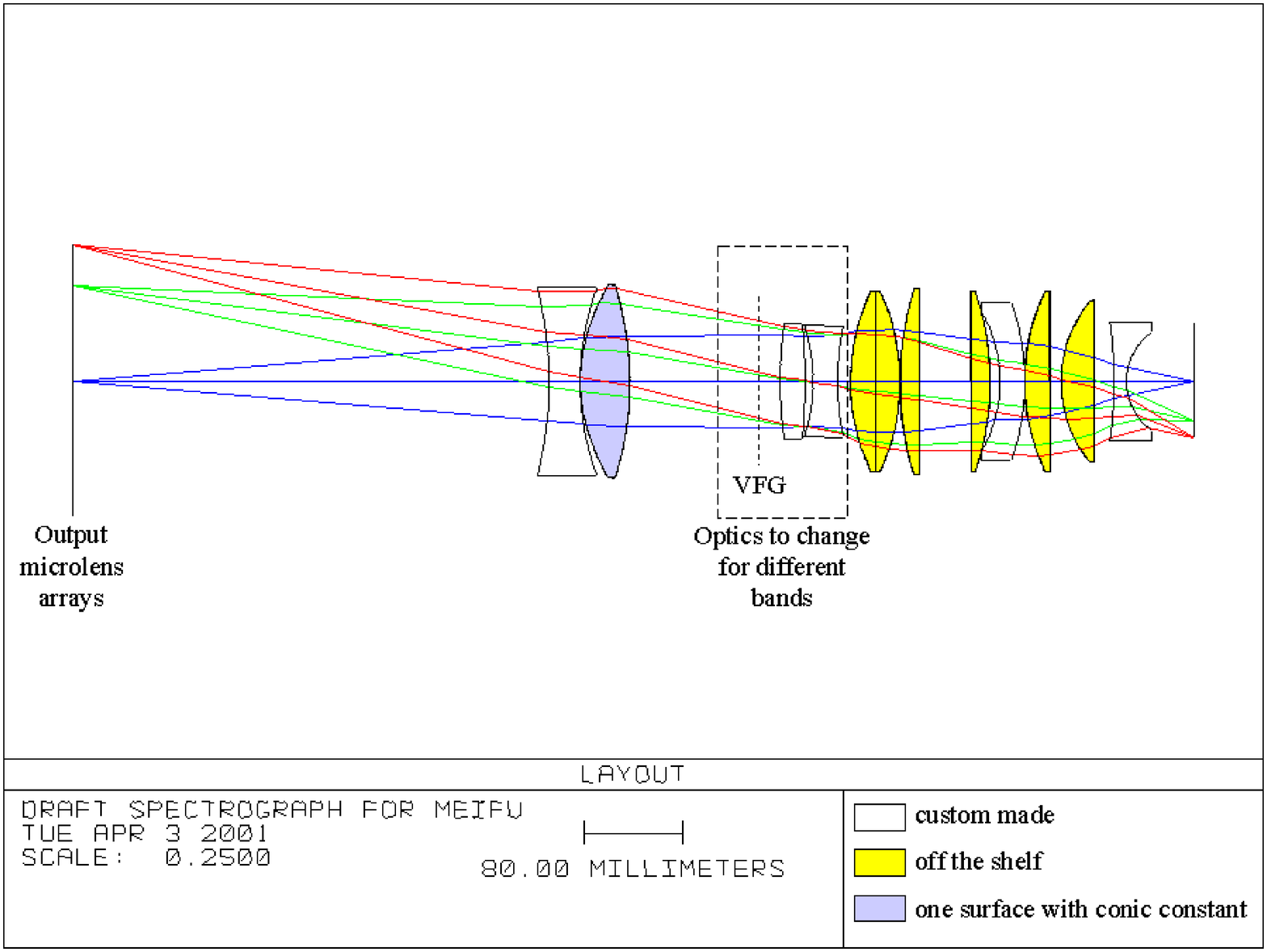}
\end{center}
\caption[]{Spectrograph concept: The design must be simple and low cost, as 27 
(12+15 identical) are required. The effective field diameter is 16 arcmin. The
camera focal ratio is ~ f/2.6. The detector arrays are 6k x 6k x 15micron 
pixels, with 50\% EE in 1 pixel; 80\% EE in 2 pixels. Note the extensive use of 
off the shelf lenses, and also the small number of optical elements that need to 
be changed in order to cover a different wavelength.}
\label{eps3}
\end{figure}

This resulted in the spectrograph concept shown in figure~\ref{eps3}.

\section{Signal-to-Noise Estimates and Predicted Numbers of Detections}

With this instrument concept, we can now make plausible estimates 
for the S/N achievable. We have tried to be conservative in these estimates, and 
to include all significant sources of noise. We list our assumptions and results 
below:
\begin{itemize}
\item Assume an exposure time of 4x8 hours ($\sim$10$^{5}$ seconds), sky 
background in V=21.8
\item  27\% sky-to-hard-disk throughput
\item  50\% of object flux in 0.6x0.6 arcsec box. (This could probably be 
improved using optimal extraction techniques, but some of that gain may be lost 
due to the objects being resolved.)
\item  All of the line flux in 2 spectral pixels (1.1 nm)
\item This leads to a 5$\sigma$ detection limit of 9x10$^{-19}$ ergs cm$^{-2}$ 
s$^{-1}$ (which is a factor 22 fainter than current narrow band surveys, see 
table~\ref{tab3}).
\item At a z of 3 (with a resulting observed $\lambda$ of 486.3 nm), this flux 
limit corresponds to a line luminosity of 7x10$^{40}$ ergs per second.
\item We would need 4x4 nights to survey the entire 420-940 nm wavelength range 
(2.45$<$z$<$6.73), for the full 5x5 arcminute field, at this depth.
\end{itemize}

\begin{figure}[htbp]
\begin{center}
\includegraphics[width=.9\textwidth]{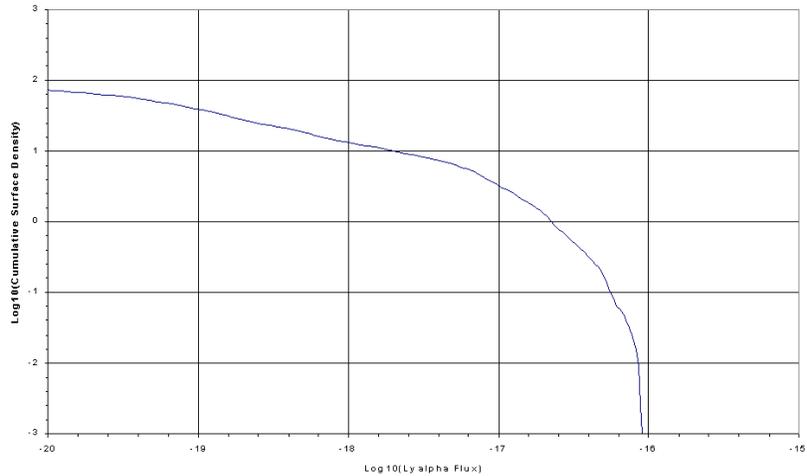}
\end{center}
\caption[]{Semi-analytic surface density prediction (having tuned the escape 
fraction to normalize to the observed counts at 3x10$^{-17}$ ergs per square cm 
per second). X-axis: Log10 (Lyman $\alpha$ flux in ergs per square cm per 
second). Y-axis: Log10 (Number per square arcmin per deltaz=1).}
\label{eps4}
\end{figure}

Now we would like to estimate roughly how many objects we might expect to detect 
in such a campaign. Since we are going substantially deeper than current data, 
we will have to rely on theoretical models to extrapolate down. For this we have 
used a semi-analytic model for galaxy formation (see [1] and references 
therein). These models predict a star formation rate (SFR), for each galaxy in 
the simulation, as a function of time. At the moment we are at liberty to tune 
the Lyman $\alpha$ escape fraction at will. We have chosen to assume that this 
is constant with redshift (probably conservative in that it is possible that the 
escape fraction may have been higher in the past due to lower metallicity of the 
objects). We then adjust the escape fraction to match the observed counts at the 
current flux limit around 2x10$^{-17}$ ergs cm$^{-2}$ s$^{-1}$, and see what is 
predicted as we lower that flux limit. The counts do to start to flatten off, as 
can be seen in figure~\ref{eps4}, but nevertheless lead to a prediction of a 
factor 12 more objects per square arcminute.

The same models predict that the surface density of objects above some 
flux limit will fall fairly rapidly with redshift. As is shown in 
table~\ref{tab3}, the current data do not strongly constrain this, but hint 
that the observed counts do not fall with redshift. Based on this we consider 
two cases - a pessimistic one where the numbers of object do indeed fall with 
redshift as predicted by the models, and an optimistic one where the numbers 
stay constant with redshift. It seems likely that the real situation is 
bracketed by these assumptions.

\begin{table}[htbp]
\caption{Lyman $\alpha$ emitter surface densities from the literature}
\begin{center}
\renewcommand{\arraystretch}{1.4}
\setlength\tabcolsep{5pt}
\begin{tabular}{lllll}
\hline\noalign{\smallskip}
z	& flux lim (1) 	& Surface Density (2)	& reference	& year \\
\noalign{\smallskip}
\hline
\noalign{\smallskip}				
2.42	& 3.00E-17	& 1.3	& Stiavelli et al.	& 2001 \\
2.81	& 3.70E-17	& 1.0	& Warren \& Moller	& 1996 \\
3.04	& 1.10E-17	& 2.7	& Moller \& Fynbo	& 2001 \\
3.09	& 3.00E-17	& 2.3	& Steidel et al.	& 2000 \\
3.15	& 2.00E-17	& 4.2	& Kudritzki et al.	& 2000 \\
3.43	& 3.00E-17	& 2.7	& Cowie \& Hu		& 1998 \\
3.43	& 1.50E-17	& 3.7	& Cowie \& Hu		& 1998 \\
3.43	& -		& 6.9	& Cowie \& Hu		& 1988 \\
4.5	& 2.00E-17	& 1.1	& Rhoads et al.		& 2001 \\
5	& -		& 2.3	& Dawson et al.		& 2001 \\
\hline
\end{tabular}
\end{center}
\noindent
(1) Fluxes in ergs cm$^{-2}$ s$^{-1}$
\noindent
(2) Surface densities per square arcminute per $\Delta$z of 1. 
\noindent
Measurements of known overdensities were reduced by a factor 6 following Steidel 
et al. (2000).
\label{tab3}
\end{table}

\begin{table}[htbp]
\caption{Predicted Numbers of detections in a 16 night exposure}
\begin{center}
\renewcommand{\arraystretch}{1.4}
\setlength\tabcolsep{5pt}
\begin{tabular}{lrr}
\hline\noalign{\smallskip}
 Z & Pessimistic & Optimistic\\
\noalign{\smallskip}
\hline
\noalign{\smallskip}
 2.46-3.20     &    773    &      773 \\
 3.20-4.10     &    497    &      993 \\
 4.10-5.17     &    303    &     1214 \\
 5.17-6.73     &    207    &     1656 \\
               &           &          \\
Total          &   1780    &     4636 \\
\hline
\end{tabular}
\end{center}
\label{tab4}
\end{table}

\begin{figure}[htbp]
\begin{center}
\includegraphics[width=.9\textwidth]{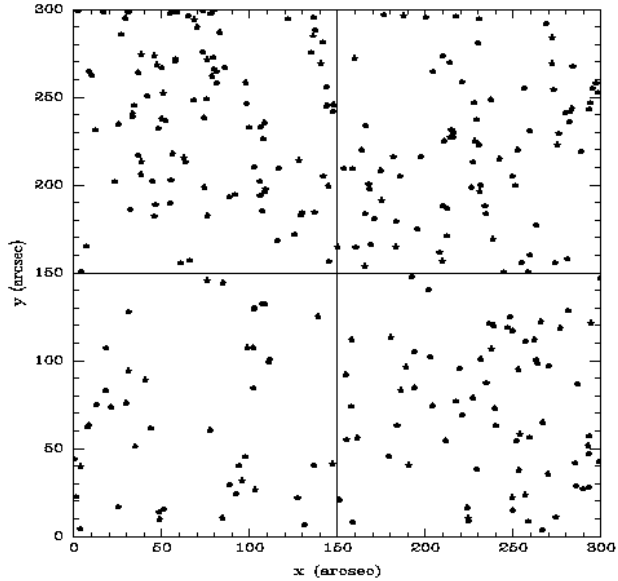}
\end{center}
\caption[]{Predicted surface density and angular distribution of 5-$\sigma$ 
detections in Lyman 
$\alpha$ from a 4 night exposure with MEIFU. Redshift ranges covered in each 
quadrant are: top-left z=2.46-3.20, top-right z=3.20-4.10, bottom-right 
z=4.10-5.17, bottom-left z=5.17-6.73. The proposed experiment would then rotate 
the spectrograph FOV by 90, 180 and 270 degrees to obtain complete wavelength 
coverage for the whole FOV in 16 nights.}
\label{eps5}
\end{figure}

From this, we can derive the number of objects we might expect to get in 16 
nights. This is given in table~\ref{tab4}. We also show in figure~\ref{eps5} 
what is predicted for the spatial distribution of the objects within a single (4 
night) exposure.

\section{Conclusions}

With MEIFU we can both push to fainter luminosities and also cover a 
much larger continuous range in redshift than has been possible with narrow band 
surveys. Within the same data cube we will see:
\begin{itemize}
\item  Lyman alpha emitters (2.5$<$z$<$6.75)
\item  {[OII]} emitters (0.13$<$z$<$1.52)
\item  faint H alpha emitters (0$<$z$<$0.43)
\end{itemize}

Another science area for which this instrument will be ideal is mapping nearby 
galaxies, including:
\begin{itemize}
\item  Starbursts (such as Messier 82)
\item  Measuring abundance gradients (e.g. comparing barred vs. unbarred 
galaxies)
\item  Looking at the stellar populations in local group dwarves (stellar 
spectral type and crude metallicities).
\end{itemize}

A few other candidate science cases that immediately suggest themselves are:
\begin{itemize}
\item  A search for intergalactic planetary nebulae.
\item  Study of high velocity knots in galactic supernova remnants (searching 
for abundance anomalies).
\item  Study of nearby planetary nebulae (measuring variations in physical 
conditions across the nebula).
\end{itemize}

In analyses of this kind, the hardest benefit to quantify is serendipity. 
However, this is also, 
history teaches us, where the greatest strides forward occur. With its massive 
data collection capability and simultaneous coverage of volumes of space, we 
believe MEIFU will be a serendipity factory. Mining 
the deep MEIFU data cubes will probably turn up many exciting and interesting 
surprises.

We would like to acknowledge considerable help and advice from R. Bacon and the 
other members of the SAURON collaboration.

%

\end{document}